\documentclass[conference]{IEEEtran}
\IEEEoverridecommandlockouts

\usepackage{cite}
\usepackage{amsmath,amssymb,amsfonts}
\usepackage{listings}
\usepackage{algorithmic}
\usepackage{graphicx}
\usepackage{booktabs}
\usepackage{textcomp}
\usepackage{xcolor}
\usepackage{todonotes}
\usepackage{booktabs}
\usepackage{tabularx}
\usepackage{url}
\usepackage{makecell} 
\usepackage{url} 
\AtBeginDocument{
  \definecolor{pdfurlcolor}{rgb}{0,0,0.6}
  \definecolor{pdfcitecolor}{rgb}{0,0.6,0}
  \definecolor{pdflinkcolor}{rgb}{0.6,0,0}
  \definecolor{light}{gray}{.85}
  \definecolor{vlight}{gray}{.95}
}
\usepackage[colorlinks=true,citecolor=pdfcitecolor,urlcolor=pdfurlcolor,linkcolor=pdflinkcolor,pdfborder={0 0 0}]{hyperref}
\usepackage{orcidlink}
\usepackage{censor}
\def\BibTeX{{\rm B\kern-.05em{\sc i\kern-.025em b}\kern-.08em
    T\kern-.1667em\lower.7ex\hbox{E}\kern-.125emX}}
\begin{document}
\newcommand{\linebreakand}{%
  \end{@IEEEauthorhalign}
  \hfill\mbox{}\par
  \vspace{-.7\baselineskip}
  \mbox{}\hfill\begin{@IEEEauthorhalign}
}

\title{Workflow Cards: Structured Summaries of Workflow Executions Using Provenance Data

\thanks{This manuscript has been authored in part by UT-Battelle, LLC, under contract DE-AC05-00OR22725 with the US Department of Energy (DOE). The publisher, by accepting the article for publication, acknowledges that the U.S. Government retains a non-exclusive, paid up, irrevocable, world-wide license to publish or reproduce the published form of the manuscript, or allow others to do so, for U.S. Government purposes. The DOE will provide public access to these results in accordance with the DOE Public Access Plan (http://energy.gov/downloads/doe-public-access-plan).}
}

\author{
      \IEEEauthorblockN{
  Nicola Giuseppe Marchioro\textsuperscript{+}\IEEEauthorrefmark{1}\,\orcidlink{0009-0005-1669-4451}
  \quad
  Gabriele Padovani\textsuperscript{+}\IEEEauthorrefmark{1}\,\orcidlink{0009-0002-3711-7643}
  \quad
  Amal Gueroudji\IEEEauthorrefmark{2}\,\orcidlink{0009-0004-4830-3139}
  \quad
  Rafael Ferreira da Silva\IEEEauthorrefmark{3}\,\orcidlink{0000-0002-1720-0928}
  \\
  Wesley Brewer\IEEEauthorrefmark{3}\,\orcidlink{0000-0002-3639-3956}
  \quad
  Valentine Anantharaj\IEEEauthorrefmark{3}\,\orcidlink{0000-0002-9356-1311}
  \quad
  Sandro Fiore\IEEEauthorrefmark{1}\,\orcidlink{0000-0002-8430-6087}
  \quad
  Renan Souza\IEEEauthorrefmark{3}\,\orcidlink{0000-0002-1794-808X}
  }
  \IEEEauthorblockA{
  \IEEEauthorrefmark{1}University of Trento, Trento, Italy\\
  \IEEEauthorrefmark{2}Argonne National Laboratory, Lemont, IL, USA\\
  \IEEEauthorrefmark{3}Oak Ridge National Laboratory, Oak Ridge, TN, USA
  }
  \thanks{\textsuperscript{+}These authors contributed equally to this work. Work done while at ORNL.}
}
\maketitle

\begin{abstract}
Model Cards and Data Cards have demonstrated the value of structured, human-readable documentation for machine learning artifacts, capturing their context, parameters, limitations, and intended use. However, these practices remain focused on static artifacts (the datasets and trained models themselves) while overlooking the workflow executions that produce, transform, and evaluate them. Such executions hold critical details about data preparation, parameter choice, runtime behavior, resource use, and intermediate transformations, precisely where bias, performance variation, and reproducibility gaps tend to originate.
To close this gap, we introduce Workflow Cards: structured summaries that condense the machine-readable provenance data of a workflow execution into a form both humans and large language models (LLMs) can read and analyze. This paper has two main parts.
First, it defines a Workflow Card template informed by a representative set of provenance questions that surface from the execution-level data missing from Model and Data Cards. Second, it evaluates how effectively LLMs use Workflow Cards to understand workflow executions compared with querying provenance databases through a schema-based interface. 
Results show that Workflow Cards provide execution-level information absent from existing card types, such as Model Cards and Data Cards, thereby filling an important documentation gap; and that Workflow Cards nearly double answer quality compared with schema-based querying, consistently across LLM-as-a-Judge and human assessments.
\end{abstract}


\section{Introduction}
\label{sec:intro}

Modern workflows generate increasingly large amounts of provenance data describing execution steps, input and output artifacts, computational environments and infrastructure, workflow parameters, Key Performance Indicators (KPIs), and interactions between  components~\cite{simmhan2005survey, souza2020workflow, prov_characterization_2024}. This information plays a fundamental role in supporting reproducibility, traceability, auditing, and workflow-level transparency~\cite{missier2014pdiff, oliveira2018analytics}, concerns closely aligned with the broader effort to treat computational workflows as FAIR scientific assets~\cite{wilkinson2025fair}.
 
Several provenance-aware systems and workflow systems have been developed to record detailed execution traces and data lineage (i.e., data provenance) in workflow runs~\cite{deelman2015pegasus, oinn2004taverna, schlegel2023mlflow2prov, souza2023towards}. These systems commonly expose the captured provenance through graph-based representations, structured databases, or machine-readable serialization formats such as those defined by the W3C PROV family of specifications~\cite{missier2013provfamily}. Although such representations provide extensive traceability, extracting meaningful workflow-level information from them typically requires specialized query tooling, familiarity with graph traversal techniques, and a deep understanding of the underlying provenance schema, barriers that substantially limit the practical accessibility of provenance information~\cite{pimentel2019survey}.
 
A parallel line of research has introduced human-readable artifacts such as \textit{Model Cards}~\cite{mitchell2019model} and \textit{Data Cards}~\cite{pushkarna2022data} to document AI components~\cite{gebru2021datasheets,
arnold2019factsheets, yang2024navigating, huggingface2022guidebook}.
However, these frameworks focus on static artifacts and offer limited insight into the executions that produce them, even though many critical provenance questions arise at runtime, when users must understand how outputs were derived, what infrastructure was employed, and how parameter choices influenced the result~\cite{simmhan2005survey, cheney2009provenance, buneman2001why, souza2020workflow}. The captured provenance holds these answers, but reaching them has required those mentioned specialized query tooling. 
Recent works explore LLM agents that translate natural-language questions into structured provenance queries~\cite{souza_llm_agents_works_sc25}; this interactive querying improves accessibility, yet it still requires the LLM agents to locate and assemble the relevant records on demand, leaving a gap for a compact, directly readable summary of the same provenance that humans and LLMs can consume.

To address this gap, we introduce \textit{Workflow Cards}, whose primary objective is to lower the  barrier to accessing workflow provenance for both human users and LLMs. Each card is a structured documentation artifact that represents workflow execution provenance through a question-oriented template. 
The template is derived from recurring provenance questions drawn from prior work~\cite{souza2020workflow, souza2025provagent, souza_llm_agents_works_sc25, SCHLEGEL2025102495, pina2025dlprov,  belhajjame2025inmemoryindexingqueryingprovenance, 6274027, 10.1145/3543873.3587559} 
and from our experiments, which reveal the fields that contribute workflow-level provenance beyond what other card types, such as Model and Data Cards, already capture. This question-oriented approach follows user-centered design, where framing outputs around user questions improves interpretability~\cite{liao2020questioning}.
A Workflow Card surfaces a workflow execution's data flow provenance, execution context, infrastructure, workflow activities, and generated artifacts.
Because LLMs excel at document-based question answering~\cite{brown2020language}, the same card serves as ready context for answering provenance questions. 
Workflow Cards complement rather than replace provenance traces and databases: they can be generated automatically from provenance systems yet remain independent of any storage backend, and may be produced via manual curation or metadata extraction pipelines.

The contributions of this work are as follows. We (i) introduce Workflow Cards as workflow-oriented documentation artifacts for exposing workflow provenance in a human- and LLM-readable format.
We (ii) organize common workflow provenance questions from the literature into three categories (input/dataset, workflow, output/model) that guide the Workflow Card design.
 We (iii) implement an open-source provenance-aware Workflow Card generation approach integrated with the provenance systems yProv4ML~\cite{PADOVANI2025102298,yprov_repo} and Flowcept~\cite{souza2023towards,flowcept_repo}. 
We (iv) evaluate Workflow Cards both as complementary documentation artifacts alongside Model and Data Cards and as an LLM-facing provenance representation, showing on a real HPC ML workflow that Workflow Cards nearly double answer quality for LLM-based provenance question answering compared with schema-based querying.

\section{Related Work}
\label{sec:rel}

\subsection{Documentation Artifacts for AI Systems}

Structured documentation artifacts significantly improve AI system transparency. Frameworks such as \textit{Datasheets for Datasets}~\cite{gebru2021datasheets}, \textit{Model Cards}~\cite{mitchell2019model}, \textit{Data Cards}~\cite{pushkarna2022data}, and \textit{FactSheets}~\cite{arnold2019factsheets} establish standard reporting practices for data collection, model behavior, and service pipelines. Recent efforts explore deeper interaction through \textit{Interactive Model Cards}~\cite{crisan2022interactive} and qualitative natural-language summaries via \textit{Report Cards}~\cite{yangreport}. These artifacts enjoy broad adoption as lightweight metadata interfaces, evidenced by analyses of thousands of repositories on platforms such as Hugging Face~\cite{liang2024s}. However, their focus remains primarily on static inputs and outputs, leaving the execution workflows connecting them largely undocumented.


\subsection{Capturing, Querying, and Representing Provenance}

Provenance systems track data lineage, parameters, and component interactions to ensure reproducibility and traceability~\cite{simmhan2005survey, freire2008survey, pimentel2019survey}. Workflow frameworks like Taverna~\cite{oinn2004taverna}, Pegasus~\cite{deelman2015pegasus}, and provenance-aware ML tooling~\cite{schlegel2023mlflow2prov, padovani2025provenance} automatically record these traces, often utilizing the W3C PROV specifications~\cite{missier2013provfamily} for interoperability. Building on these foundations, researchers have formalized provenance querying to extract derivation relationships and outcomes~\cite{buneman2001why, cheney2009provenance}, expanding recently into machine learning lifecycles~\cite{schlegel2022survey,souza2020workflow} and agentic workflows~\cite{souza2025provagent,federated_2026,controla_2025}.

Although provenance-aware systems provide extensive traceability capabilities, prior surveys consistently identify accessibility and usability as major limitations of existing provenance representations~\cite{pimentel2019survey}, since provenance is typically exposed through graph structures, database schemas, or low-level serialization formats optimized for storage and machine querying rather than for human interpretation. 
Consequently, while provenance systems preserve workflow information faithfully, they rarely offer lightweight interfaces for communicating provenance semantics to human users or LLM-based systems.
A complementary line of work packages an entire workflow run together with its provenance. The Workflow Run RO-Crate profile~\cite{rocrate_workflow_run} standardizes 
machine-readable, FAIR packaging of workflow executions. 
Workflow Cards are orthogonal to, not a replacement for, such formats: rather than exhaustively packaging a run, a Workflow Card extracts its provenance into a compact, query-driven summary intended for direct human reading and LLM consumption.
This positioning distinguishes Workflow Cards from the broader ecosystem of machine-oriented metadata schemas for ML assets, such as Croissant~\cite{croissant} for datasets, FAIR4ML~\cite{fair4ml} for models, and PROV-ML~\cite{souza2020workflow} for ML provenance, which standardize structured, machine-readable descriptions. Workflow Cards are complementary to these schemas: they sit one layer above by condensing the provenance data that such schemas help represent into a compact, human- and LLM-readable summary of workflow executions, and can coexist with them~\cite{gustafsson2025workflowhub}.



\section{Methodology}
\label{sec:method}

We propose Workflow Cards to summarize workflow execution provenance data in a lightweight, interpretable format, and evaluate them across two independent benchmarks sharing the same protocol. Benchmark~I assesses whether Workflow Cards
provide execution-level information absent from existing Model and Data Cards. Benchmark~II evaluates their accessibility advantage for LLMs compared to a lightweight schema-based interface to a provenance database.
Two LLMs act as judges throughout: a smaller model, \textit{nemotron-nano-3}~\cite{nvidia2025nemotron3nanoopen} ($\sim$30B parameters), and a substantially larger one, \textit{gpt-oss-120b}~\cite{openai2025gptoss120bgptoss20bmodel}($\sim$120B parameters).
We use an LLM-as-a-Judge approach because it scales evaluation across the full question set and has been shown to correlate strongly with human judgment on answer-rating tasks~\cite{li2024llmsasjudgescomprehensivesurveyllmbased, zheng2023judgingllmasajudgemtbenchchatbot}, and we additionally validate it against an independent human rater in Benchmark~II. The answering LLMs, however, differ by design. In Benchmark~I, a single separate model (gpt-4o~\cite{openai2024gpt4ocard}) answers every card
configuration, so that performance differences reflect the cards rather than the model. In Benchmark~II, the two judge models also serve as answering models, spanning model scale to test whether the Workflow Card advantage holds for both a smaller and a larger model.

\subsection{Categorizing Workflow Provenance Questions}

Workflow Cards organize relevant information around common provenance access patterns rather than exposing exhaustive traces. To identify these patterns, we analyzed prior work~\cite{souza2020workflow, souza2025provagent,souza_llm_agents_works_sc25,SCHLEGEL2025102495,pina2025dlprov,belhajjame2025inmemoryindexingqueryingprovenance,6274027,10.1145/3543873.3587559} and extracted recurring questions, grouping them into three macro-categories: input (dataset),  workflow, and output (model) provenance. These categories served as a conceptual guide for the template design.
Starting from these categories, we iteratively refined the Workflow Card template and  manually designed a new question set of 33 questions spanning workflow (13), model (10), and data (10) aspects, released with our templates and reproduction code~\cite{wfcards_artifacts}. These questions examine which information is common across documentation artifacts and which information is specific to particular artifact types.
During the design process, we established three principles for the Workflow Card. First, it should expose workflow-level provenance rather than replicate metadata already documented in Model or Data Cards. For the representative use case of foundation model fine-tuning, with an input Model Card, one or more Data Cards, and an output Model Card, this means information already held by those cards, such as a detail in the pretrained Model Card, need not be repeated in the Workflow Card. Second, it should represent an immutable execution, so each Workflow Card reflects one specific run and does not evolve after generation. Metadata unrelated to the run itself, such as administrative or policy information that may change independently, is therefore excluded. Third, like the model and Data Cards, it should offer a high-level overview with the smallest footprint possible while maintaining strong question-answering performance.

\subsection{Benchmark I: Cross-Artifact Information Complementarity}

The first benchmark focused on evaluating the amount of provenance information exposed by Workflow Cards and measuring the overlap between Workflow Cards and existing documentation artifacts. While many different types of cards have been proposed in the literature, this study was restricted to machine learning workflows and focused exclusively on Model Cards and Data Cards, with one card per pipeline artifact: a pretrained Model Card, a fine-tuning Data Card, and a fine-tuned Model Card.
To evaluate informational overlap across documentation artifacts, we constructed a set of simulated machine learning fine-tuning workflows using publicly available Hugging Face repositories, referenced in Table~\ref{tab:references}.

\begin{table}[htbp]
\centering
\caption{Overview of Benchmarked Use Cases, divided between Input Model Cards, Data Cards, and Output Model Cards}
\label{tab:references}
\vspace{2mm}
\scriptsize
\setlength{\tabcolsep}{2pt}
\begin{tabularx}{\columnwidth}{>{\raggedright\arraybackslash}X >{\raggedright\arraybackslash}X >{\raggedright\arraybackslash}X}
\toprule
\textbf{Input Model} & \textbf{Data Card} & \textbf{Output Model} \\
\midrule
\texttt{meta-llama/}\newline\texttt{Llama-3.2-}\newline\texttt{3B-Instruct} & 
\texttt{avaliev/}\newline\texttt{chat\_doctor} & 
\texttt{prithivMLmods/}\newline\texttt{Llama-Doctor-3.2-}\newline\texttt{3B-Instruct} \\
\midrule
\texttt{nvidia/}\newline\texttt{Nemotron-Cascade-}\newline\texttt{2-30B-A3B} & 
\texttt{nvidia/}\newline\texttt{Nemotron-Cascade-}\newline\texttt{2-RL-data} & 
\texttt{empero-ai/}\newline\texttt{openNemo-Cascade-}\newline\texttt{2-30B-A3B} \\
\midrule
\texttt{ibm-nasa-}\newline\texttt{geospatial/}\newline\texttt{Prithvi-100M} & 
\texttt{ANI00/}\newline\texttt{Crop-Health-Monitor} & 
\texttt{ibm-nasa-geospatial/}\newline\texttt{Prithvi-100M-multi-}\newline\texttt{temporal-crop-\newline classification} \\
\midrule
\texttt{distilbert/}\newline\texttt{distilbert-base-}\newline\texttt{uncased} & 
\texttt{cybersectony/}\newline\texttt{PhishingEmail}\newline\texttt{Detectionv2.0} & 
\texttt{cybersectony/}\newline\texttt{phishing-email-\newline detection-}\newline\texttt{distilbert\_v2.4.1} \\
\midrule
\texttt{meta-llama/}\newline\texttt{Llama-3.2-3B} & 
\texttt{ai4privacy/}\newline\texttt{pii-masking-200k} & 
\texttt{chinu-codes/}\newline\texttt{llama-3.2-3b-}\newline\texttt{pii-redactor-lora} \\
\bottomrule
\end{tabularx}
\end{table}

For each workflow configuration, we collected the pretrained Model Card, associated Data Cards, and fine-tuned Model Card. Since public Workflow Cards do not yet exist, we simulated executions and generated Workflow Cards with the proposed
  template. Workflow-level fields were synthetically generated using \textit{Claude Sonnet 4.6}, constrained by the real Model and Data Cards. Benchmark~I is therefore a structural complementarity analysis: it tests whether the template can
  represent execution-level information absent from Model and Data Cards, not the factual correctness of synthetic fields. Benchmark~II complements it with real provenance captured by Flowcept and yProv4ML (Figure~\ref{fig:example_card}).

We evaluated provenance question answering under three different information settings: (i) Full Knowledge, in which all documentation artifacts were concatenated and provided simultaneously to the LLM. This configuration provides complete workflow-level context and is considered the top-level answer; (ii) Single-card, in which each card was evaluated independently by providing only that documentation artifact; (iii) Leave-one-out, where all cards except one were provided to the LLM. This configuration evaluates the informational contribution of the omitted artifact.

In the context of this benchmark, information is defined in terms of the ability to answer a question. A documentation artifact is considered informative with respect to a question if an LLM can correctly answer that question using only the information contained within the artifact itself. The prompts instructed the LLMs to avoid inferring or hallucinating. 

 The Full Knowledge configuration establishes the upper-bound informational context available for each workflow execution and was additionally double-checked, corrected, and improved with any missing information. 
This reference therefore represents the best answer obtainable from the full documentation set, curated and completed by the authors, and serves as the comparison target for the other configurations.
Generated answers were evaluated by the two LLM judges. 
Each LLM rated the answer from 0.0 to 1.0 against the reference answer, where 1.0 denotes full consistency with the reference, 0.0 an incorrect or unsupported answer, and intermediate values partial correctness.
Each LLM had its temperature set to 0.0 for the queries, and an additional statistical check was repeated five times to ensure that answer variability did not produce outliers. 
To additionally characterize the information carried by each card, we computed the pairwise cross-entropy across the documentation artifacts.

\subsection{Benchmark II: Workflow Cards vs. Schema-based Provenance Querying}

Benchmark~II evaluates whether Workflow Cards improve accessibility compared to schema-based querying of the underlying provenance data.
It uses the Workflow Card template to describe a scientific machine learning pipeline for the DLESyM~\cite{cresswell2025deep} climate forecasting model. The workflow spans four steps: (i) pre-processing raw data using the REDI~\cite{brewer2026data} tool, (ii) fine-tuning DLESyM, (iii) forecasting, and (iv) evaluating forecasts via Empirical Orthogonal Function~\cite{hannachi2007empirical}.

The workflow was instrumented using both Flowcept and yProv4ML to capture provenance  across all steps. The card generator maps captured entities, activities, agents, task timestamps, resource records, and I/O metadata into the template blocks. Fine-grained task records are grouped by workflow stage, start and end times are reduced to activity durations, status values are summarized as counts, resource telemetry is aggregated at run and activity level, and selected inputs and outputs populate the Significant Workflow Artifacts block.
  
From the benchmark question set, a 17-question subset targeting mainly workflow provenance was used, since questions outside that scope would have shifted the comparison away from information accessibility.
We compared two representations of this same captured provenance, differing only in how it is presented to the LLM. The first is a lightweight schema-based querying approach~\cite{souza_llm_agents_works_sc25}, where the LLM is only given the schema and structural provenance rather than the raw records. Data, in this case, is stored into JSONL buffers, and retrieved from a MongoDB database.  
The second is a Workflow Card generated from the same provenance data and supplied directly as context.
This comparison contrasts a pre-computed representation with on-demand retrieval: the Workflow Card packages the relevant provenance ahead of time as static context, whereas schema-based querying requires the model to navigate the interface and assemble the necessary provenance per question.
Both representations were answered by the two models and scored under the same LLM-as-a-Judge protocol.
 A human expert evaluator independently assessed the Benchmark~II answers as a calibration check for the automated judges. The human ratings are used to measure whether the LLM-as-a-Judge scores follow the same trend as manual assessment, while
  the LLM judges provide scalable scoring across the benchmark.

\section{Results}
\label{sec:results}

The results are organized into three parts. Section~\ref{sec:template} presents the final template. Section~\ref{sec:bench1} reports Benchmark~I, comparing the information carried by Dataset, Model, and Workflow Cards. Section~\ref{sec:bench2} reports Benchmark~II, comparing Workflow Cards with schema-based provenance querying.

\subsection{Workflow Card Template}
\label{sec:template}

\begin{figure*}[!h]
\centerline{\includegraphics[width=0.86\linewidth]{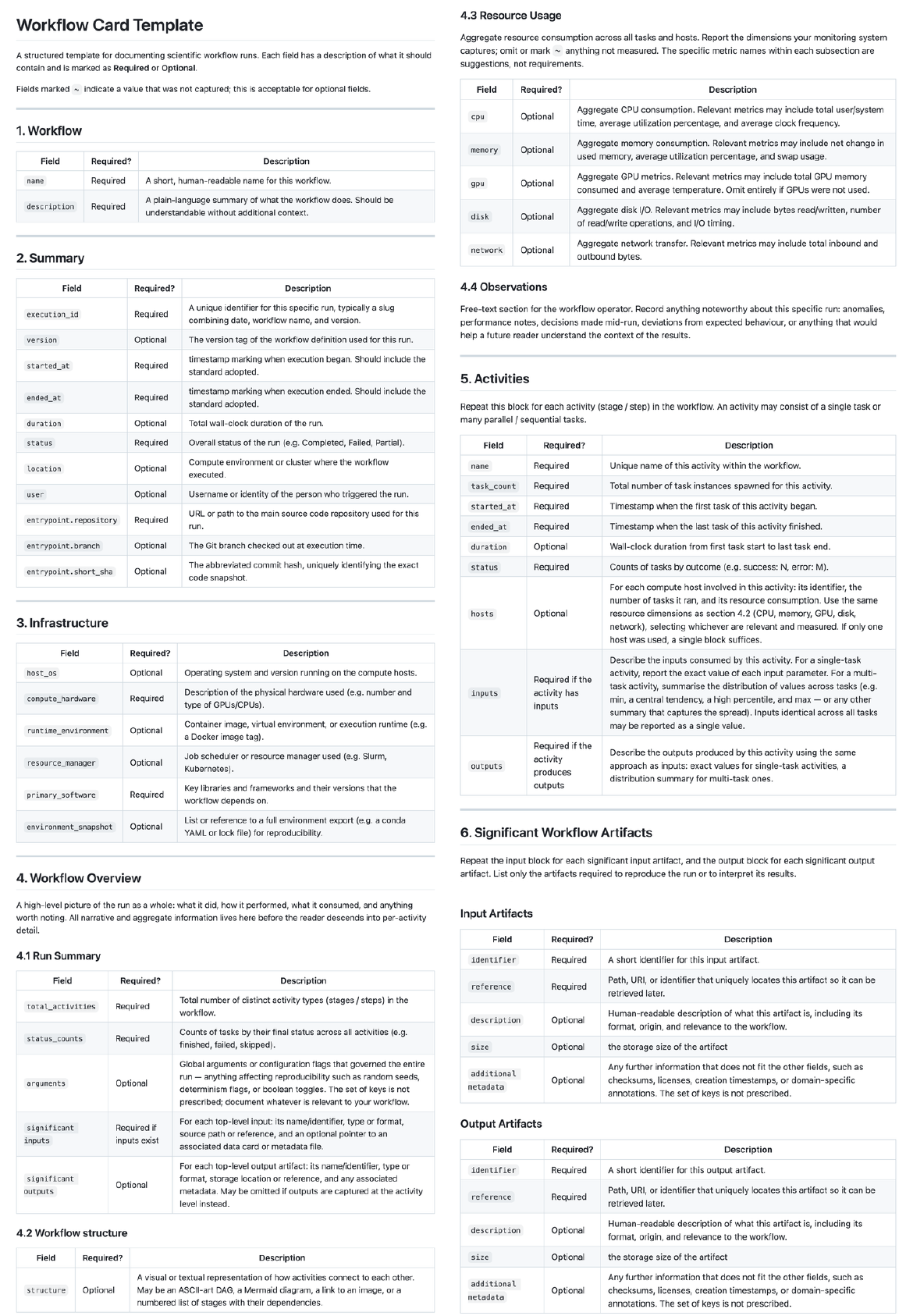}}
\caption{Template structure of a Workflow card, illustrating the key fields and layout used to represent workflow information. Available online~\cite{wfcards_artifacts}.}
\label{fig:templatever}
\end{figure*}

The template was produced through an iterative process of drafting, testing on both benchmarks, and analyzing the results. The final Workflow Card template is shown in Figure~\ref{fig:templatever} and is organized as a top-down summary that begins with a high-level view of the run and progressively descends into per-activity detail. It opens with a short Workflow block that names and describes the run in plain language, followed by a Summary block that records the identity and lifecycle of the execution, including a unique execution identifier, the workflow version, start and end timestamps, total duration, overall status, and the entrypoint repository, branch, and commit that pin the exact code used. An Infrastructure block then captures the computational environment, covering the operating system, hardware, runtime, resource manager, and a snapshot of the software dependencies, so that the run can be situated and, where possible, reproduced. The Workflow Overview gathers run-level aggregates such as activity and task counts, governing arguments, aggregate resource usage across CPU, memory, GPU, disk, and network, and a free-text Observations field for anomalies or decisions made mid-run, before the Activities block repeats a compact record for each stage that reports its tasks, timing, status, hosts, and summarized inputs and outputs. A final Significant Workflow Artifacts block lists only the inputs and outputs needed to reproduce or interpret the run, each identified by a short name and a resolvable reference. Every field is marked as Required or Optional, and a tilde denotes a value that was not captured, which keeps the card lightweight while signaling missing information rather than hiding it. Because each card describes a single immutable execution, it deliberately omits metadata that may evolve independently of the run, and it favors aggregate, high-level descriptions over exhaustive traces so that it stays small enough to read directly and to pass to an LLM as context. 

\begin{figure*}[!h]
\centerline{\includegraphics[width=\linewidth]{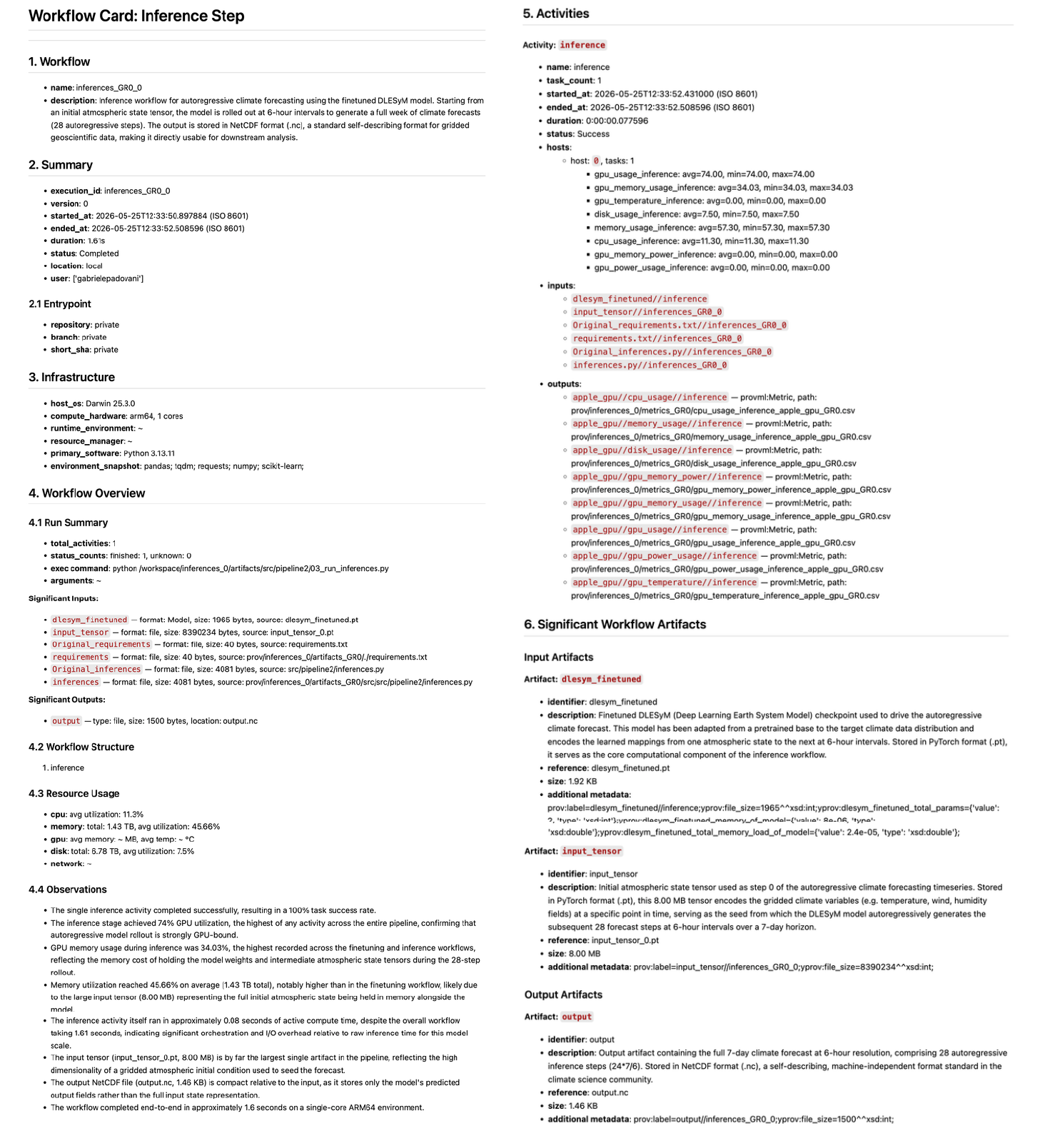}}
\caption{Example Workflow card describing an inference step, performing weather forecasting based on an initial atmospheric condition. Available online~\cite{wfcards_artifacts}.}
\label{fig:example_card}
\end{figure*}



\subsection{Benchmark I}
\label{sec:bench1}

The first benchmark quantifies the drop in performance  when only a single card is provided (the Data Card, pretrained or fine-tuned Model Card, or Workflow Card), compared to the full set of all of them concatenated. 
The least informative card, the one with the largest average drop from the all-cards reference, is the pretrained Model Card. 
The Workflow Card, by contrast, matches the full set (Figure~\ref{fig:degred}), reaching 0.619 against the all-cards reference of 0.611 while being, on average, 33.2~kB smaller, showing the same answer quality at a fraction of the context size, roughly 8k fewer tokens at about four characters per token.

In these figures, the score is calculated as the average of the LLM-as-a-Judge ratings. Variability for these two initial charts is high, as each column encodes a large number of disparate questions, repeated for five different use cases. 

\begin{figure}[htbp]
\centerline{\includegraphics[width=\linewidth]{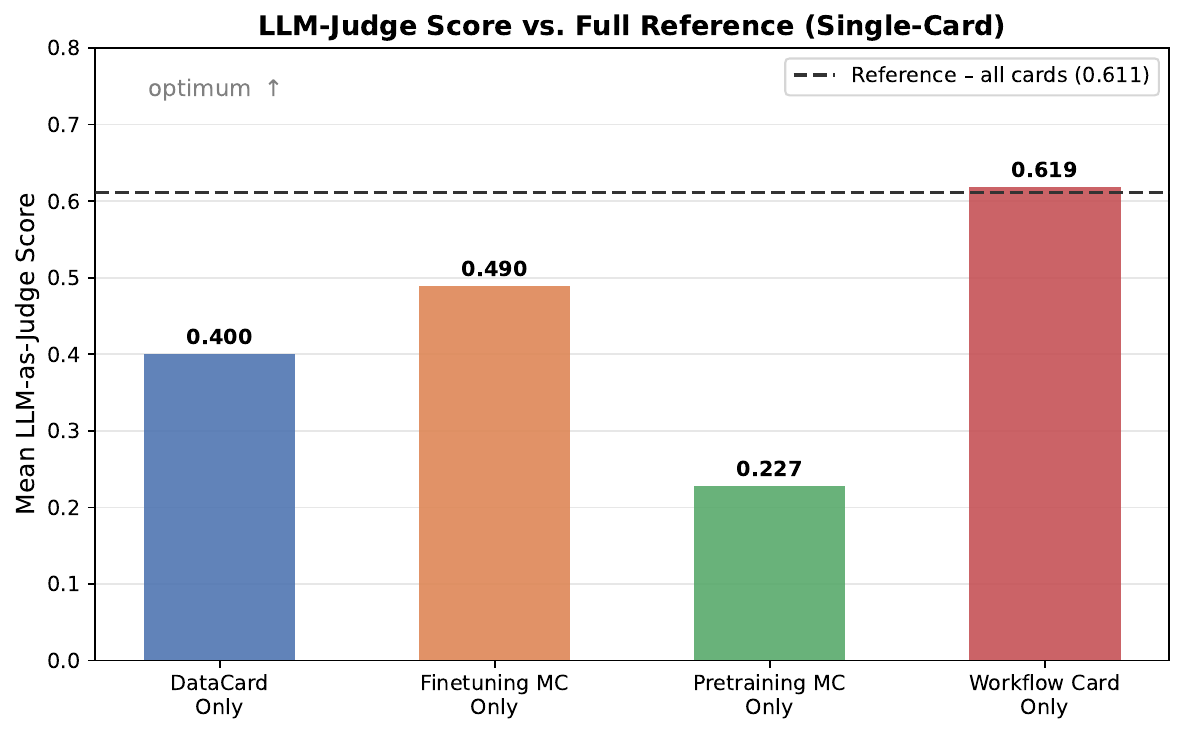}}
\caption{Degradation of the mean of LLM as judge score across different single-card contexts, compared to the all-cards reference. Higher is better. }
\label{fig:degred}
\end{figure}

Then, we tested the inverse: the drop when all cards but one are provided (Figure~\ref{fig:leave}), e.g., leaving out one of the Model, Data, or Workflow cards compared to the reference set. 
Although higher scores generally indicate better performance, in this analysis, the focus is on the impact of removing individual artifacts. A larger decrease in score when an artifact is omitted suggests information that is not easily recovered from the remaining artifacts, making it more informative.

\begin{figure}[htbp]
\centerline{\includegraphics[width=\linewidth]{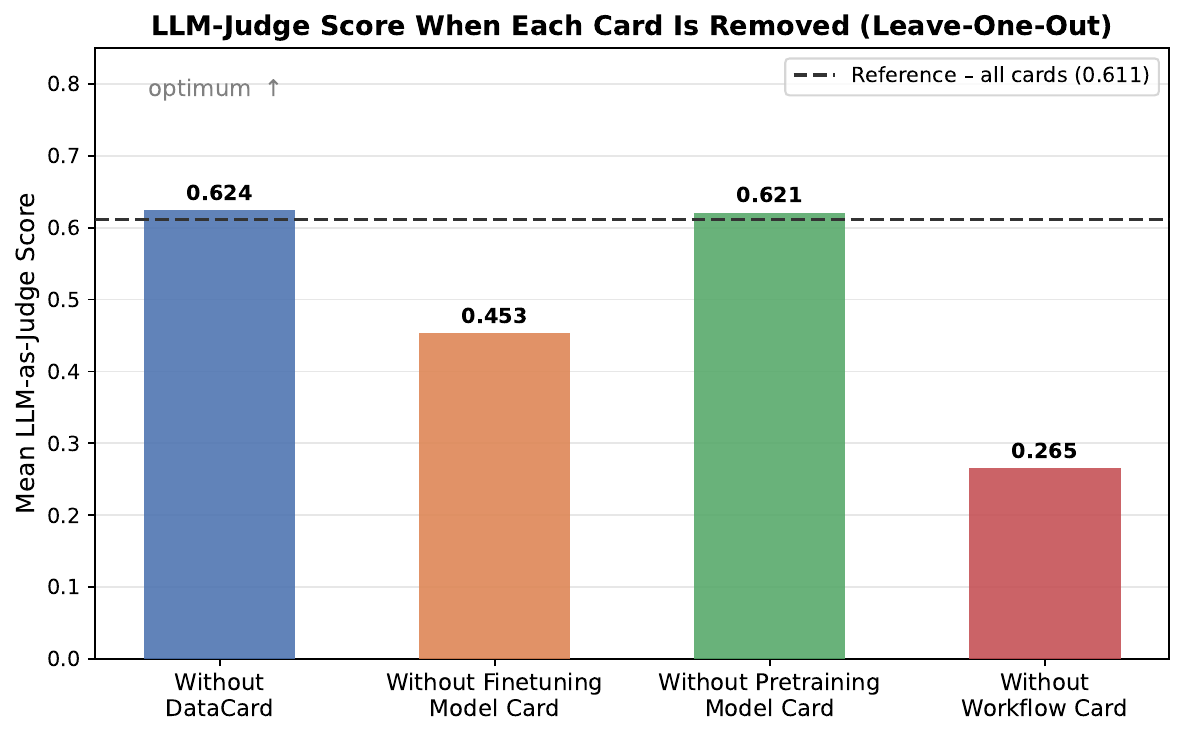}}
\caption{Leave-one-out performance analysis. Although higher scores indicate better performance, the key signal is the drop relative to the full setup.
Removing the Data Card or pretrained Model Card has minimal impact, and removing the fine-tuned Model Card a moderate one, whereas excluding the Workflow Card causes the largest decrease, indicating it contains the most unique and relevant information for answering the evaluation questions.
}
\label{fig:leave}
\end{figure}


In this leave-one-out setting, removing the Workflow Card produces the largest drop-off, followed by the Fine-tuned Model Card, which contains information about the workflow's final output.
Quantitatively, the execution-level information unique to the Workflow Card more than doubles accuracy over the remaining cards: removing it drops the mean score from the all-cards reference of $0.611$ to $0.265$ (a $2.3\times$ drop), by far the largest effect of any artifact; removing the Data Card or the Pretrained Model Card, by contrast, changes the score only marginally.
To explain these differences, we computed the pairwise cross-entropy between cards (Figure~\ref{fig:crossentropy}). It accounts for the small effect of removing the Pretrained Model Card: much of its information is also captured by the Fine-tuned Model Card, but not the reverse, an asymmetry reflected in the cross-entropy between the two cards.

\begin{figure}[htbp]
\centerline{\includegraphics[width=\linewidth]{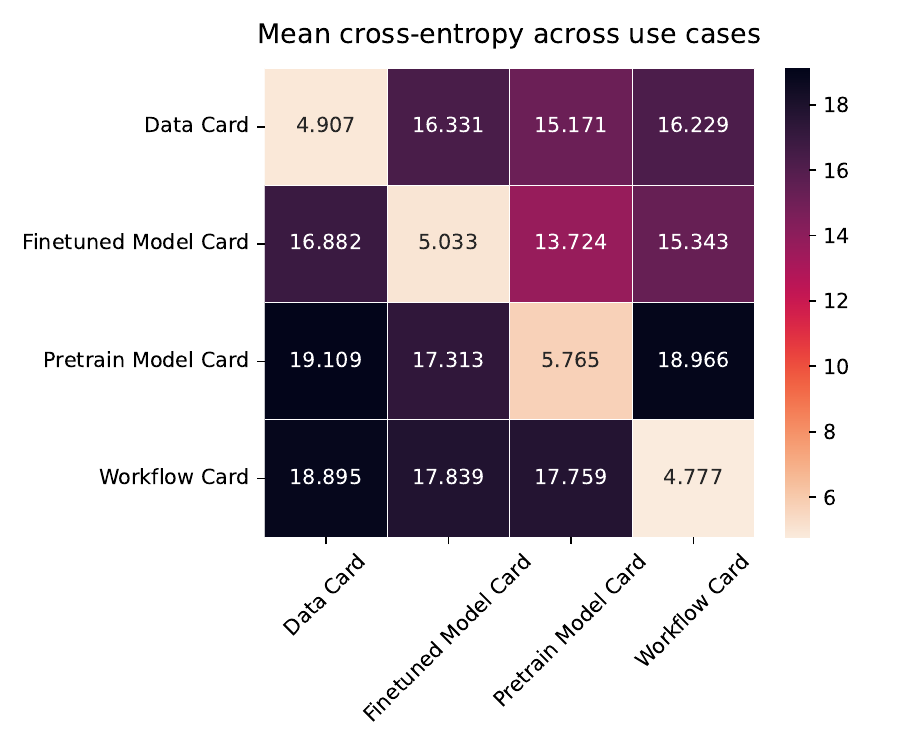}}
\caption{Matrix of mean cross-entropy scores across all use cases evaluated between different pairs of provenance card types. A lower score indicates less "surprising" information. }
\label{fig:crossentropy}
\end{figure}

Finally, we quantified how well the answering models perform when given only one artifact type and asked questions targeting a specific card type (e.g., dataset-related questions given a Data Card). Table~\ref{tab:perfocus} reports this setting; its columns are the all-cards reference (All~Cards) and the four single-card conditions: the Data Card, the fine-tuned Model Card (FT~MC), the pretrained Model Card (PT~MC), and the Workflow Card. 
 Among the Data and Model Cards, the card matching a question's focus performs best (e.g., the Data Card on data questions). 
 However, the Workflow Card alone matches or exceeds the All-Cards reference on every question type (Data, Model, and Workflow), despite using less context.
Two checks guarded the comparison. First, no artifact exceeded 25~kB, keeping card sizes comparable. Second, the models maintained their answer quality on questions a human rater judged difficult. 
The reported scores are means over the five use cases: single-card results are stable across them (standard deviations near 0.06, with the pretrained Model Card more variable at 0.13), while the leave-one-out conditions vary  more (standard deviations up to 0.27), since the effect of removing a card depends on the specific workflow.

\begin{table}[htbp]
\centering
\small 
\setlength{\tabcolsep}{4pt} 
\caption{Performance isolated per question focus, compared against the all-cards reference. Higher is better.}
\label{tab:perfocus}
\begin{tabular}{lccccc}
\toprule
\makecell[l]{\textbf{Question}\\\textbf{Focus}} & \makecell[c]{\textbf{All}\\\textbf{Cards}} & \makecell[c]{\textbf{Only}\\\textbf{DataCard}} & \makecell[c]{\textbf{Only}\\\textbf{FT MC}} & \makecell[c]{\textbf{Only}\\\textbf{PT MC}} & \makecell[c]{\textbf{Only}\\\textbf{Workflow}} \\
\midrule
Data     & 0.6612 & 0.6154 & 0.4744 & 0.2736 & 0.6884 \\
Model    & 0.6035 & 0.4190 & 0.5716 & 0.3488 & 0.6688 \\
Workflow & 0.3999 & 0.1172 & 0.2626 & 0.1088 & 0.4005 \\
\bottomrule
\end{tabular}
\end{table}


\subsection{Benchmark II}
\label{sec:bench2}


We first checked for large self-preference effects in the LLM-as-a-Judge ratings. Each model rated both systems' responses on a 0.0 to 1.0 scale, and an independent human rating was included as a calibration reference.
Figure~\ref{fig:judge} shows no clear self-preference pattern: nemotron-nano-3 rates its own answers lower than gpt-oss-120b's, while gpt-oss-120b grades slightly more generously overall. The human ratings follow the same trend as the model judges, supporting the use of the LLM judges as a calibrated automated evaluation.

\begin{figure}[!ht]
\centerline{\includegraphics[width=\linewidth]{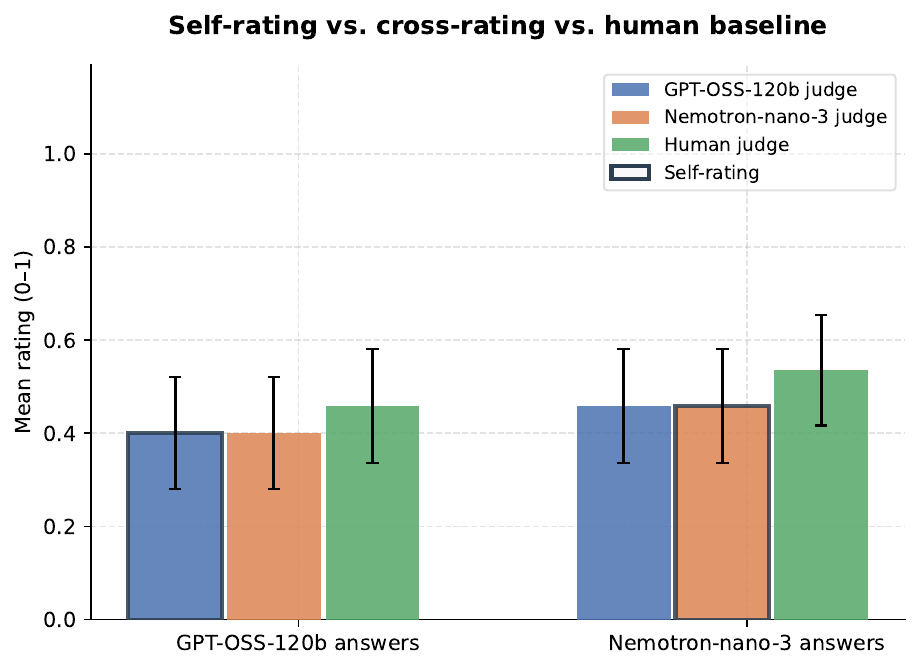}}
\caption{Comparison of mean ratings between gpt-oss-120b answers and nemotron-nano-3 answers across different judges and human baselines.}
\label{fig:judge}
\end{figure}

\begin{figure}[!ht]
\centerline{\includegraphics[width=\linewidth]{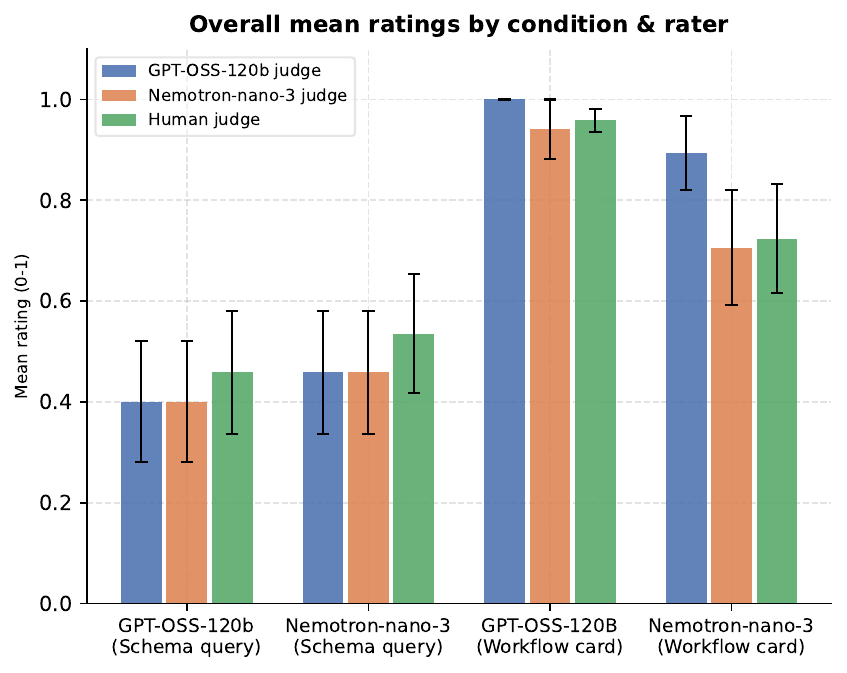}}
\caption{Overall mean ratings for gpt-oss-120b and nemotron-nano-3 responses under the two provenance representations (schema-based querying vs.\ Workflow Card), as evaluated by both models with the LLM-as-a-Judge framework and by human raters.}
\label{fig:perf}
\end{figure}

We then compared question answering under schema-based querying and under Workflow Cards. 
On average, the card setting is noticeably better (Figure~\ref{fig:perf}): across both answering models, moving from schema-based querying to Workflow Cards nearly doubles the mean rating (from $\approx$0.45 to $\approx$0.87), an improvement on which the LLM-as-a-Judge and human evaluations agree. The spread of ratings is similar for both.
Agreement among judges is strongest when cards are the input. For the schema-based approach, the LLM judges score considerably harsher than the human rater, likely because raw schema-and-structure responses are harder for an LLM to interpret than a pre-generated card.


\section{Discussion, Future Work, and Artifact Availability}

 Our results show that pre-summarizing provenance into a card substantially eases LLM consumption compared to navigating raw schemas. Because Workflow Cards are themselves generated from the provenance captured by systems such as Flowcept,
the two representations are complementary rather than competing: a provenance agent can emit a Workflow Card and use it as context, or consume one to answer queries. Integrating on-demand card generation into interactive provenance agents is a promising direction for future work.

Our study has limitations. Benchmark~I evaluates the informational structure of Workflow Cards on synthetically generated content rather than factual provenance, while Benchmark~II complements it by validating on an instrumented HPC workflow.
Benchmark~II, in turn, covers a single workflow assessed by one human evaluator, and because the LLM judges also produced the answers they rate, the human ratings
  (Fig.~\ref{fig:judge}) serve to validate the automated judgments. 
  Quantitatively, the human ratings agree closely with the LLM judges (Pearson $r=0.74$ for the mean of the two judges, and $0.66$ and $0.77$ individually; mean absolute  difference $0.13$ on the 0 to 1 scale).
Finally, both benchmarks target machine-learning workflows: fine-tuning pipelines and one climate-ML workflow. The template is designed around execution concepts shared by many workflow systems, including run identity, infrastructure,
  activities, timing, status, parameters, and input/output artifacts. In non-ML domains such as genomics, computational fluid dynamics, or weather simulation, the Significant Workflow Artifacts block would document domain outputs such as FASTA,
  HDF5, Zarr, or NetCDF files rather than model weights. Validating these adaptations on non-ML scientific workflows remains future work.

To support reproducibility, the Workflow Card template, benchmark question sets, LLM prompts, and the code to reproduce the benchmarks are publicly available~\cite{wfcards_artifacts}.
 Additionally, the updated versions of Flowcept~\cite{flowcept_repo} and
yProv4ML~\cite{yprov_repo} featuring Workflow Card generation support are open-source.

\section{Conclusion}

Understanding how artifacts were created is critical for identifying biases, preventing misuse, and improving overall explainability. We introduced Workflow Cards as interpretable documents of execution processes connecting datasets and models, filling a vital gap in current reporting standards. Our evaluation demonstrates that Workflow Cards capture unique, process-level provenance absent from existing artifacts and, on a real HPC workflow, nearly double LLM answer quality on provenance questions compared with schema-based querying. By making execution provenance directly readable to both people and LLMs, this work takes one more step toward the transparency and reproducibility of scientific workflows.

\tiny
\section*{Acknowledgment}

\small
This work was partially funded under the National Recovery and
Resilience Plan (NRRP), Mission 4 Component 2 Investment 1.4 - Call for tender No. 1031 of 17/06/2022 of Italian Ministry for University and Research funded by the European Union – NextGenerationEU (proj. nr. CN\_00000013), and the RI-SCALE project (Grant Agreement 101188168). 
This research used resources of the Oak Ridge Leadership Computing Facility at the Oak Ridge National Laboratory, supported by the Office of Science of the U.S. Department of Energy under Contract No. DE-AC05-00OR22725.
This material is based upon work supported by the U.S. Department of Energy (DOE), Office of Science, Office of Advanced Scientific Computing Research, under Contract DE-AC02-06CH11357.
The authors used LLMs (Claude Sonnet 4.6 and GPT-5.5) to assist with grammar, sentence structure, and length control across sections. As described in the paper, an LLM was additionally used to generate synthetic metadata for the Benchmark I. These LLMs produced no scientific analyses, interpretations, or conclusions.

\bibliographystyle{./IEEEtran}
\bibliography{./IEEEabrv,./refs}


\end{document}